\documentclass[conference]{IEEEtran}
\IEEEoverridecommandlockouts
\usepackage{cite}
\usepackage{amsmath,amssymb,amsfonts}
\usepackage{algorithmic}
\usepackage{graphicx}
\usepackage{textcomp}
\usepackage{xcolor}
\usepackage{booktabs}  
\usepackage{pifont}    
\usepackage{url} 

\newcommand{\tick}{\ding{51}}   
\newcommand{\cross}{\ding{55}}  
\def\BibTeX{{\rm B\kern-.05em{\sc i\kern-.025em b}\kern-.08em
    T\kern-.1667em\lower.7ex\hbox{E}\kern-.125emX}}
\begin{document}

\title{GATMesh: Clock Mesh Timing Analysis using Graph Neural Networks}
\author{\IEEEauthorblockN{ 
Muhammad Hadir Khan,
Matthew R. Guthaus }\\
\IEEEauthorblockA{Computer Science and Engineering\\University of California Santa Cruz, Santa Cruz, CA 95064\\ 
\{mkhan33, mrg\}@ucsc.edu}}
\newcommand{\todo}[1]{{\bf TODO: {#1}}\\}
\maketitle

\begin{abstract}
Clock meshes are essential in high-performance VLSI systems for minimizing skew and handling PVT variations, but analyzing them is difficult due to reconvergent paths, multi-source driving, and input mesh buffer skew. SPICE simulations are accurate but slow; yet simplified models miss key effects like slew and input skew. We propose GATMesh, a Graph Neural Network (GNN)-based framework that models the clock mesh as a graph with augmented structural and physical features. Trained on SPICE data, GATMesh achieves high accuracy with average delay error of 5.27ps on unseen benchmarks, while achieving speed-ups of 47146x over multi-threaded SPICE simulation.
\end{abstract}


\section{Introduction}

\begin{figure}[htb]
    \centering
    \includegraphics[width=0.5\linewidth]{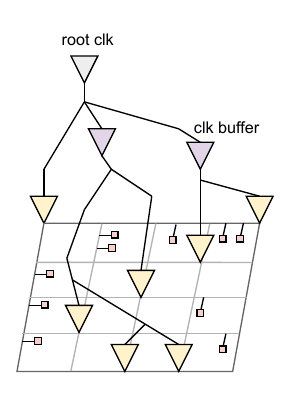}
    \caption{A clock tree driving the root clock to the mesh buffers (yellow) through clock tree buffers (purple). Mesh buffers then drive the mesh grid with sinks (red squares) to improve skew due to re-converging paths.
    \label{fig:mesh}}
    \vspace{-0.4cm}
\end{figure}

Clock distribution networks (CDNs) are fundamental to high-performance VLSI
designs, ensuring precise synchronization across millions of sequential
elements. Among various architectures, clock meshes stand out for their
superior robustness against process, voltage, and temperature (PVT) variations,
offering low clock skew and improved tolerance to uncertainties in modern
fabrication processes. By leveraging multiple redundant paths as shown in Figure~\ref{fig:mesh}, clock meshes
effectively distribute the clock signal across the chip. However, these
advantages come with significant challenges.

One of the primary difficulties in analyzing clock meshes arises from the
simultaneous multi-source driving of the network. These sources originate from
an underlying clock tree, where variations in delay among different branches
introduce varying input arrival times to the mesh. Such variations,
combined with the inherently re-convergent paths within the mesh, make
traditional modeling techniques ineffective. SPICE-based simulations, though
highly accurate, are computationally prohibitive for large designs, while
simplified models, such as first-order delay~\cite{desai:meshsizing} are inaccurate and fail to capture essential
effects like slew propagation, which critically impacts setup and hold
constraints. Higher-order model reduction techniques, such as Arnoldi-based
methods, are more computationally expensive and often struggle to converge due to the complex interdependencies within the mesh.

Furthermore, integrating SPICE-based solutions with static timing analysis
(STA) is not practical due to the significant runtime overhead and the
difficulty of incorporating SPICE-level accuracy into conventional STA tools.
In practice, designers of the highest-performance chips often rely on manual
simulation~\cite{power4_clock, power6_clock, desai:meshsizing, amd:piledriver, amd:bulldozer, guthaus:highperf}, clock mesh tuning, and over-design of the mesh to meet timing
requirements. This over-design results in increased power consumption and the
disconnect between simulation and STA poses a major limitation in timing
verification workflows, restricting efficient design exploration and
optimization.

To address these challenges, we propose a novel approach using Graph Neural
Networks (GNNs) trained on SPICE simulation data to predict clock delays and
slew rates efficiently. By treating the clock mesh as a graph, the GNN learns
to approximate complex timing relationships, significantly reducing
computational costs while maintaining high accuracy. This data-driven approach can then be
applied to unseen designs and enables a rapid, scalable alternative for performance estimation.
 
We demonstrate that our GNN-based framework can provide accurate delay and slew predictions, offering a
practical solution for integrating clock mesh analysis into modern VLSI design
flows. Our approach lays the groundwork for AI-driven design automation in
clock network synthesis, bridging the gap between accuracy and efficiency in
timing analysis. Specifically, this paper:
\begin{itemize}
    \item Proposes the first GNN methodology for clock-mesh timing analysis.
    \item Demonstrates the effectiveness of GNNs in predicting both clock delays and signal slew for better control of setup/hold times.
    \item Validates the proposed approach against SPICE simulations and prior clock-mesh analysis techniques on a range of unseen open-source benchmark designs.
\end{itemize}

\section{Background}

\subsection{Clock Meshes}

Clock meshes have been extensively adopted in high-performance microprocessors by leading semiconductor 
companies~\cite{power4_clock, power6_clock, desai:meshsizing, amd:piledriver, amd:bulldozer, guthaus:highperf}. These
companies rely on clock meshes to achieve minimal skew, enhance timing
reliability, and improve tolerance to process, voltage, and temperature (PVT)
variations. However, the
significant area and power overhead associated with traditional clock meshes
have limited their widespread adoption in general ASIC design, necessitating research into more
efficient implementations.

Several methodologies
have been proposed to address the power and area challenges of clock meshes. MeshWorks~\cite{meshworks} introduced an automated
framework for clock mesh synthesis, enabling efficient planning, optimization,
and tradeoff balancing between skew and power consumption. It integrates mesh
planning and optimization algorithms, which reduce buffer area, wire length, and
power. MeshWorks also introduced mesh reduction
techniques, which selectively remove redundant mesh segments while maintaining
variation tolerance. This approach significantly reduces power and resource
consumption without degrading performance. There have been other works to
optimize clock meshes~\cite{teng:mesh, wilke:buffer, guthaus:mesh, friedman:mesh}, which have further improved upon these results. All of these results, however, have typically relied on restricted models of delay or heuristics instead of delay modeling. In addition, none of the approaches have considered clock signal slew which can have a large impact on timing performance.


\subsection{Graph Neural Networks}
Deep learning over grid structured data like images has shown promising results using Convolutional Neural Networks (CNNs). These same CNNs, however, have been problematic on irregular structured data like graphs because of graph isomorphism (i.e. same graphs with different node orderings). Graph Neural Networks (GNNs), however, have been a promising technique to solve this by using localized message passing between neighboring nodes and combining the resulting features into graph, node, or edge predictions. 

GNNs struggle with deep layers because of the over-smoothing problem \cite{oversmoothingsurvey} where after multiple steps of the aggregation operation the node features become indistinguishable from each other. This is particularly problematic on graphs with large spans since communication between distant nodes is limited. Transformers are a recent approach to aid such global communication within a single layer~\cite{transformer}, but scale poorly for large graphs. Jumping Knowledge (JK) connections~\cite{jknet} are another approach to dynamically aggregate node representations from both different layers of a GNN to capture both local and global information. While JK connections have somewhat different function than transformers, they scale much better for large graphs. They also enable the model to adaptively choose the most relevant neighborhood range for each node from multiple layers, improving performance on tasks with diverse structural requirements.

\subsection{GNNs and Clocks}

While GNNs have been used for many aspects of EDA, they have had limited
use on clock trees and buffer insertion. BufFormer~\cite{bufformer} uses GNNs to predict buffer insertion
topologies and locations on non-clock nets. There has also been work on using
GNNs to predict the timing of nets, but not necessarily clock
nets~\cite{Ye2023:WireTiming, Cheng2020:WireTiming}. More importantly, these prior works did not focus on nets with multiple drivers and reconvergent networks like clock meshes. Single nets and
buffered trees are well adapted to traditional STA and higher-moment RC-network analysis
methods, but the challenges of clock meshes are not addressed by these methods. 


\section{GATMesh}
Clock meshes are regular, but they have complexities due to
non-uniform drivers, differing input arrival times, irregularity due to reduced meshes and
blockages, and uneven distribution of clock sink capacitance. These
graph characteristics, however, can be used as input features on GNNs to model
clock mesh timing. In particular, we propose to use GNNs inductively to infer results on designs that it has not previously seen.

GNNs extend neural networks to graph-structured data, making them ideal for
modeling clock meshes, where nodes represent mesh points and edges capture
connectivity. Unlike CNNs, which operate on regular grids, GNNs process
general graphs, allowing them to learn timing relationships in the meshes more
effectively. Additionally, GNNs handle isomorphism better than CNNs, ensuring
consistent predictions across equivalent, but reordered, mesh structures. However,
GNNs for clock meshes are still challenged by the over-smoothing problem with deep layers.

\begin{figure}[tb]
    \centering
    \includegraphics[width=1.0\linewidth]{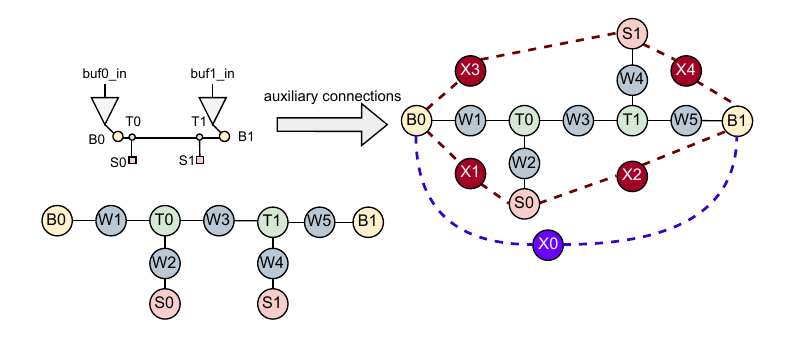}
    \caption{\textbf{Left:} A single mesh wire, its buffers and its graph representation. \textbf{Right:} Graph representation after adding buffer contention ($X0$) and sink driving ($X1-X4$) auxiliary connections.}
    \label{fig:aux-conn-graph}
\end{figure}

We model the SPICE netlist of clock meshes as an undirected graph as shown in Figure~\ref{fig:aux-conn-graph}. The SPICE netlist clock mesh is a resistor-capacitor (RC) network driven by mesh buffers. We model the interconnect as Wire nodes (W) in the graph. The clock sink pins are modeled as load capacitance to ground which we represent as Sink nodes (S). The mesh buffers are Buffer nodes (B) that connect directly to mesh wires.  Each clock sink is connected to a mesh wire through a ``stub'' wire represented by an additional W node and a Tap node (T).  The exemplary graph representation of a single mesh wire with two sinks and two buffers is shown in Figure~\ref{fig:aux-conn-graph}. 

One challenge with the above model, and many GNN models in general, is that the number of hops in the graph between buffers and sinks depend on the mesh implementation and the resulting graph topology. If there are many sinks connected to a mesh wire, the mesh wire gets divided into smaller segments with additional T and W nodes. In Figure~\ref{fig:aux-conn-graph}, for example, the buffer $B0$ drives both the sinks $S0$ and $S1$ with $S0$ being four hops away, whereas $S1$ is six hops away because it must pass through the intermediate $T0$ and $T1$ nodes and the divided $W1$ and $W3$ node resulting from the split mesh wire. Any GNN model would require an increasingly deeper number of layers to allow message passing through deep graphs, yet so many deep layers would also result in well-known over-smoothing of features in connected regions~\cite{oversmoothingsurvey}.  

While it is possible to reduce the number of nodes by using edge features, most GNN models are primarily \textit{node centric}. We tried edge feature representations and saw inferior results. We observe, however, that edge feature representations are only a constant factor shallower than a node representation so the problems with deep GNN models still apply in most clock mesh designs.  Instead, to enable deep message passing, we rely on two complementary techniques: a static, design inspired augmentation of the graph in this section and a learnable model using JK Connections in Section~\ref{sec:gnn-arch}. The static augmentation of the graph leverages insight from clock mesh designers and previous clock mesh generation papers while the learnable connections leverage machine learning to capture higher-dimension information based on the graph and its corresponding model architecture.

\subsection{Sink Driver Auxiliary Connections}

The key feature of clock meshes is that there are redundant mesh drivers that drive the mesh, and therefore, every sink. While all mesh buffers are driving the same mesh, a clock sink will primarily be driven by nearby clock mesh buffers, but not exclusively. Similar intuition on this locality was used in some prior heuristic algorithms. For example, some clock mesh synthesis algorithms utilize sink coverage metrics to decide where to place buffers~\cite{guthaus:mesh}. In practice, we have found that considering the two nearest buffers for every sink adds appropriate insight into the neighborhood of buffers driving a sink. This makes sense because each sink is on a mesh wire and buffers are likely coming from two directions along the mesh wire. We include an auxiliary connect between the two nearest buffers in terms of path driving resistance (i.e., the buffer output resistance plus the shortest interconnect path) and corresponding capacitance. For example, in Figure~\ref{fig:aux-conn-graph}, both $S0$ and $S1$ are driven by the nearest buffers $B0$ and $B1$ so auxiliary connections are added along with nodes $X1$-$X4$ to model the features associated with the connections. 

\subsection{Buffer Contention Auxiliary Connections}

While the above auxiliary connections model the shared driving of a sink by multiple buffers, buffers can also impede switching if they have drastically different mesh buffer input offset delays, differing output signal slew, and buffer gate delay due to uneven sink density or mismatch between the local capacitance and buffer size. This was first observed in Wilke et al. which analyzed the impact of input tree skew on mesh power~\cite{wilke:buffer} but used some novel circuit schemes to reduce the short circuit power. To consider this effect, we add auxiliary connections between mesh buffers so that message passing can learn whether they are switching at the same time or have skew which may affect the nearby sink delays. In practice, we have found that adding the four nearest buffers to every buffer adds appropriate insight into the neighborhood of buffers.  While we used two buffers for clock sinks because they are on a mesh wire with two directions, a mesh buffer is usually at an intersection of four mesh wires, so considering four buffers possibly corresponds to all four cardinal directions. Again, like the sink driver connection, we model the minimum path driving resistance and capacitance. For example, in Figure~\ref{fig:aux-conn-graph}, $B0$ and $B1$ can have different driving strengths and input delay/slew from the root clock tree, so they are connected via an auxiliary connection node $X0$ to model the features associated with the connection.

\begin{table}[ht]
\caption{Features and Labels (Rows) on Node Types (Columns)}
\centering
\begin{tabular}{lccccc}
\toprule
& Buffer  & Tap & Wire  & Sink  & Aux  \\
\textbf{Features} & (\textbf{B}) & (\textbf{T}) & (\textbf{W}) & (\textbf{S}) & (\textbf{X}) \\
\midrule
input\_delay (ps)        & \tick  & \cross  & \cross  & \cross & \cross \\
input\_slew (ps)         & \tick  & \cross  & \cross & \cross  & \cross  \\
cap (fF)         & \tick  & \cross  & \tick  & \tick  & \tick \\
res (ohms)         & \tick  & \cross & \tick  & \cross & \tick  \\
total\_res (ohms)   & \tick & \tick  & \cross  & \tick & \cross  \\
min\_res (ohms)     & \tick & \tick  & \cross  & \tick & \cross  \\
region\_cap (fF)  & \tick  & \tick & \cross  & \tick & \cross  \\
xy\_loc ($\mu$m)     & \tick  & \tick  & \cross  & \tick  & \cross \\
\midrule 
& Buffer  & Tap & Wire  & Sink  & Aux  \\
\textbf{Labels} & (\textbf{B}) & (\textbf{T}) & (\textbf{W}) & (\textbf{S}) & (\textbf{X}) \\
\midrule
delay\_out (ps) & \cross  & \cross  & \cross  & \tick & \cross \\
slew\_out (ps) & \cross  & \cross  & \cross  & \tick & \cross \\
\bottomrule
\end{tabular}

\label{tab:node-features}
\end{table}

\subsection{Input Features}

The node features used for the GNN model are shown in Table~\ref{tab:node-features}. The check mark indicates non-zero values for these features whereas the cross mark indicates $0$ for these features.

The $input\_delay$ and $input\_slew$ are the delay and slew respectively of the inputs arriving at the mesh buffers after propagating through a top-level tree (see Figure~\ref{fig:mesh}). Since the top-level tree is constructed using a zero-skew Elmore delay model for interconnect and buffers, there is generally some skew between the mesh buffer inputs. In addition, each mesh buffer has differing input slew which can affect the delay of the mesh buffer. 

The capacitance ($cap$) and resistance ($res$) features depend on the origin and type of graph node. For buffers, $cap$ is the buffer output capacitance and $res$ is the buffer output (driving) resistance. For wire segments such as mesh wires or stub wires (i.e., from sink to mesh wire), the $res$ and $cap$ depend on the routing layer parasitics and interconnect length. For sink nodes, the $cap$ feature is the load capacitance of the input pin of the sequential element.  The $xy\_loc$ are the x and y coordinates of the nodes in the layout. 

Besides the direct features, we also add some pre-computed features to help the model learn about signal propagation in meshes. In particular, we compute the $region\_cap$ which is the total capacitance within a particular resistance from a node. This effectively models the nearby capacitance to a node and is similar to the ``downstream'' capacitance in an Elmore delay model. We chose the resistance limit as the resistance of a single mesh wire so that it will consider all of the nearby sinks.

In addition to capacitance, we also model an aggregate driving resistance to a node from the mesh buffers. The $total\_res$ is the path resistance (i.e. summation of the conductances) of the minimum resistance path from each buffer to the node. The $min\_res$ is the single lowest resistance from any buffer to the node. The $total\_res$ provides insight into how many buffers are simultaneously driving a given node while $min\_res$ provides insight into how well a node is driven by any one buffer. 

While the augmented features provide insight into the resistance and capacitance behavior of a driven node, they do not help with the topology and message passing. This is done through the auxiliary connections for buffer contention and sink driving which also have $res$ and $cap$ obtained from the minimum resistance paths.

\subsection{Output Labels}

\begin{figure}[htb]
    \centering
    \includegraphics[width=1.0\linewidth]{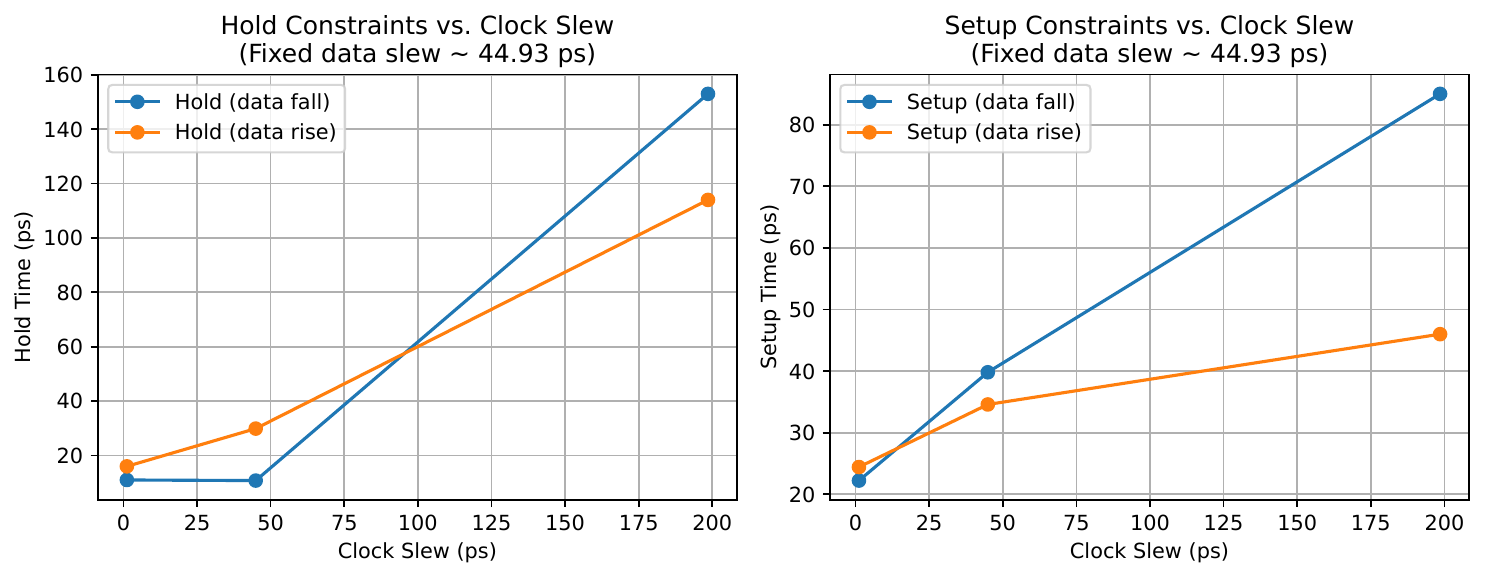}
    \caption{Clock signal input slew has a significant impact on setup and hold times and needs to be considered as a primary objective of clock synthesis.}
    \label{fig:slew-impact}
\end{figure}

The sink nodes have output labels for the $delay$ and $slew$ computed by SPICE simulation. While $delay$ is important for skew, the $slew$ is equally important as it can affect the setup and hold times of sequential elements. While we could obtain better delay accuracy by not considering slew, slew has a significant impact on the setup and hold times of sequential elements. In Nangate45, for example, a DFF\_X1 flip-flop has setup and hold constraints as shown in Figure~\ref{fig:slew-impact}, assuming a single fixed data input slew. Most prior clock synthesis results do not consider clock signal slew, but this is an important metric in high-performance clock synthesis and is therefore an important part of our model output.

\subsection{Model Architecture}
\label{sec:gnn-arch}
\begin{figure*}[t]
    \centering
    \includegraphics[width=1.0\textwidth]{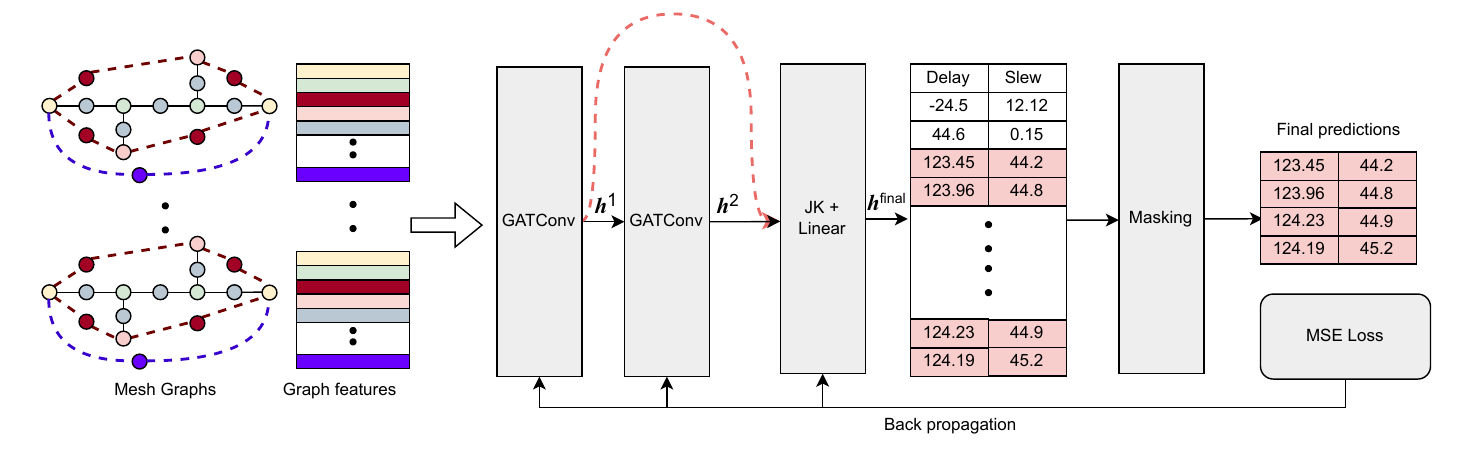}
    \caption{Model architecture overview. As an example three GATConv layers are shown with Jumping Knowledge connections (shown in red). Masking is performed to only calculate the loss for predictions on sink nodes. }
    \label{fig:model}
\end{figure*}

Our model architecture, illustrated in Figure~\ref{fig:model} with only three layers, uses the Graph Attention Network (GAT) convolution layers~\cite{gat} with JK connections~\cite{jknet}. We predict both delay and slew for clock sink nodes so we have a node-level, continuous regression problem. Furthermore, since we do not want to need values for all the nodes, we use a mask to compute the loss only from sink nodes. 

The mesh graphs along with node features from Table~\ref{tab:node-features} are input to GAT Convolutional Layers (GATConv). At each layer $k$, new node features $h^{k+1}$ are computed by aggregating the information from the neighbors according to the attention mechanism. Multiple heads are simultaneously computed with learnable attention weights which allows us to give importance to the different neighbors when aggregating their features. For example with four heads, GAT assigns four different weight vectors to the neighbor features when aggregating and concatenates the weighted features together. Intuitively, this may correspond to multiple combinations of driving wires or buffers from different directions on the mesh. 

To compute the final node features $h^{final}$ we utilize the JK connections which are  shown in Figure~\ref{fig:model} with the red dashed line. Generally, the features from each layer $h^k, h^{k+1}, ... h^l$ where $l$ are the number of layers are passed to the final layer where they are either concatenated, max-pooled or passed to a Bi-directional LSTM. In our case, we use the max-pooling where it learns to select the most informative layer for each feature coordinate. Concatenation is the most straightforward but it only optimizes the weights to combine the features in a way that works best for the overall dataset which can cause underfitting when the graphs are too complex. On the other end, LSTM is more complex and can cause overfitting. Of these aggregation schemes, JK connections with max pooling shows better performance overall in deep GNNs~\cite{jknet} like ours.

The JK connections serve a different purpose compared to the auxiliary connections we added to our graphs. They provide learnable combination of features in each neighborhood of the RC mesh whereas the aux connections model two physical aspects of a clock mesh: multiple buffer contention and multiple sink drivers which are statically chosen.

\section{Methodology}

\subsection{Training Data Generation}

We analyzed some of the open-source benchmark designs with OpenROAD~\cite{openroad} and implemented them using the default flow in the Nangate45 technology. We  used the design statistics of area, number of sinks, and sink density to generate various synthetic designs for clock synthesis as shown in Table~\ref{table:data-stats}.  Clock meshes were generated for these synthetic designs using both uniform buffering~\cite{meshworks} and non-uniform buffering~\cite{guthaus:mesh}. The uniform and non-uniform buffered meshes use the same Meshworks mesh sizing with minimum total wire length while the ``fixed'' mesh used a fixed $50\mu m$ mesh with non-uniform buffering. We also observed large designs in OpenROAD having clustered sinks with whitespace blockages for macros and memories. To simulate that we also create clusters of sinks for larger synthetic designs to reflect macro blockages. 

Top-level trees were created to drive the clock buffers using the well known DME~\cite{boese:zeroskew,chao:zeroskew} combined with balanced buffering~\cite{tsai:clocksizing}. The top-level trees are zero-skew according to the Elmore model but may have some skew due to mismatch between SPICE and Elmore delay models. The arrival input delay and slew of the tree leaves are used as features of the mesh buffers as shown in Table~\ref{tab:node-features}.

\begin{table}[htb]
\centering
\caption{Statistics of the training dataset}
\label{table:data-stats}
\setlength\tabcolsep{3pt}
\begin{tabular}{|c|r|r|r|r|r|r|}
\hline
& \multicolumn{3}{|c|}{Number of Mesh Types} & Area  & Density   \\
\cline{2-4}
&  Uniform & Non-Uni. & Fixed & ($\mu m^2$) & (sinks/$\mu m^2$) \\
\hline
Small & 100 & 100 & 100 & $[500,1500]$ & $0.033$ \\
\hline
Medium & 100 & 100 & 100 &  $[40000,60000]$ & $[0.01, 0.04]$ \\
\hline
Large & 100 & 100 & 100 &  $[180000,220000]$ & $0.021$  \\

\hline
\end{tabular}
\end{table}

\subsection{Training Procedure}

Our model makes delay and slew predictions for all the nodes but we compute the loss only for the sink nodes using a mask with a Mean-Squared Error (MSE) loss function together with the target labels which are computed by running SPICE simulation on the clock meshes. The gradients are calculated and back-propagated to the JK layer as well as to the preceding GATConv layers. This allows the JK layer to learn the importance of each feature for different granularity of neighborhood and relate the features to the target delay and slew. Simultaneously, the GATConv layers learn to assign appropriate attention weights to the neighbors at the current level for the delay and slew predictions.

We used ADAM~\cite{adam-opt} as an optimizer with a learning rate of $7.5e^{-4}$ and weight decay of $5e^{-4}$ for L2-regularization. For non-linear activations we used ELU~\cite{elu}. We used batch size of $1$ and train with a default $1000$ epochs with random shuffling of training data each epoch. We also implemented early-stopping in our training loop with a patience of $20$. With an ample amount of synthetic training data, L2-regularization, and early-stopping, we did not find the need to add dropout to the layers since the model does not overfit.

\subsection{Model Parameters}
The model hyperparameters we used are shown in Table~\ref{table:hyperparameters}. Through empirical evaluations, we found better performance by having a deep model architecture and therefore used eight convolution layers. This is due to the depth of the mesh graph model -- a sink that is driven by a closest buffer will be at minimum four hops away as shown in Figure~\ref{fig:aux-conn-graph}. However, there can be many other sinks nearby and each of them would create tap point nodes on the mesh wire thereby increasing the depth of the graph. We tried other convolution layer types besides GATConv, but found that learning the weights to combine neighbors is very important, likely because the importance is related to the resistance features of neighbor wires.  The auxiliary connections add the insight of nearby driving buffers and improve message passing; however, these only model the minimum resistance paths and not the full mesh, so the local connections through RC mesh nodes also generate information on how the sinks interact within their local interconnect. We observed that four attention heads performed well, likely due to the four orthogonal directions at each mesh point.

\begin{table}[ht]
\centering
\def\arraystretch{1.5}
\caption{Model Hyperparameters}
\label{table:hyperparameters}
\setlength\tabcolsep{6pt}
\begin{tabular}{|c|c|}
\hline
Parameter & Value \\
\hline
\multicolumn{1}{|l|}{Convolution layer} & GATConv \\
\hline
\multicolumn{1}{|l|}{Number of convolution layers}         & 8 \\
\hline
\multicolumn{1}{|l|}{Number of convolution channels}    & 64 \\
\hline
\multicolumn{1}{|l|}{Number of attention heads}            & 4 \\
\hline
\multicolumn{1}{|l|}{JK Aggregation}                & max \\
\hline
\end{tabular}
\end{table}

\subsection{Experimental Setup}

We implement GATMesh in Pytorch~\cite{pytorch} with the Pytorch Geometric library~\cite{pytorch_geometric}. The Ngspice and first-order analysis as well as the benchmark generation was run on a dual AMD EPYC 7542 32-Core Processor machine with 512Gb of memory. The training and testing of GATMesh was done on an  NVIDIA GeForce RTX 4090 24GiB card. 
   
\section{Results}

\begin{table*}[htb] 
\caption{Timing analysis on open-source designs compared to Ngspice simulation}
\begin{center}
\begin{tabular}{|c|r|r|r|r|r|r|r|r|} 
\hline
\textbf{Open-Source}&\multicolumn{4}{|c|}{\textbf{Stats}}&\multicolumn{2}{|c|}{\textbf{Delay error (ps)}}&\multicolumn{2}{|c|}{\textbf{Slew error (ps)}}\\
\cline{2-9} 
\textbf{Designs} & \textbf{\textit{Area ($\mu m^2$)}} & \textbf{\textit{Sinks}} & 
\textbf{\textit{Mesh Size}} & \textbf{\textit{Buffers}} & \textbf{\textit{GAT-Mesh}} & \textbf{\cite{desai:meshsizing}} & \textbf{\textit{GAT-Mesh}} & \textbf{\cite{desai:meshsizing}} \\
\hline
gcd & 1082.88 & 35 & 2x2 & 4 & 1.76 &  79.63 & 0.31 & N/A\\
\cline{1-9} 
aes & 52785.04 & 562 & 6x6 & 31 & 2.65 &  157.34 & 2.69 & N/A\\
\cline{1-9} 
ibex & 57585.27 & 1932 & 15x15 & 168 & 2.91 &  203.39 & 7.48 & N/A\\
\cline{1-9} 
dynamic\_node & 54950.01 & 2257 & 15x15 & 180 & 3.58 &  199.78 & 3.18 & N/A\\
\cline{1-9} 
tiny\_rocket & 193770.36 & 4035 & 18x18 & 268 & 3.67 & 278.45 & 7.74 & N/A\\
\cline{1-9} 
jpeg & 198612.62 & 4383 & 19x19 & 287 & 3.46 &  266.13 & 5.10 & N/A\\
\hline
swerv & 387381.12 & 11180 & 34x34 & 799 & 7.99 & 197.13  & 7.87 & N/A\\
\hline
swerv\_wrapper & 1056663.92 & 11214 & 35x35 & 538 & 16.16 & 219.71  & 13.50 & N/A\\
\hline
\cline{1-9}
\textbf{Average Error} & & & &  & \textbf{5.27} & \textbf{200.19} & \textbf{5.98} & \textbf{N/A} \\
\cline{1-7}
\hline
\end{tabular} \label{table:delay-slew-results}
\end{center}
\end{table*}

We compare our GATMesh model with both SPICE which is the ``golden'' model and a well-known first-order approximation~\cite{desai:meshsizing} that is used in most of the prior clock mesh synthesis research papers. While RC network analysis may provide more detail than first-order delay estimates, it still falls far short of SPICE-level accuracy because it typically neglects key effects in clock meshes. These methods often assume linear superposition of buffers driving the mesh and ignore the nonlinear behavior of buffers, particularly when multiple drivers with differing input arrival times and output contention are present. These methods also oversimplify the effects of the redundant mesh paths, where signals from the different drivers interact. These interactions can potentially reinforce or cancel each other. As a result, such models can misestimate both delay and slew, making them only marginally more accurate than first-order approximations but still significantly less reliable than full SPICE simulations. We also attempted to use some higher order models, but these always failed to converge. The first-order models, on the other hand, would usually converge. Our work focuses on comparing with SPICE and a first-order model as a representative of the other heuristic models.

\subsection{Training and Model Accuracy}

We train on a dataset of $900$ synthetic designs shown in Table~\ref{table:data-stats}. The training/development/test set split is $80/10/10$, so in total we have $720$ samples for training, $90$ for validation and $90$ for testing. The training takes $\leq$ $200$ epochs to converge in under $2$ hours. We train and validate using Mean Squared Error (MSE) loss on delay and slew of the sink nodes. The squared term helps the model to penalize the mispredictions especially when the predictions are far from the true labels and is convex. 
The training converges with an MSE loss of $23.87$ps, validation set MSE loss of $31.83$ps, and test set MSE loss of $33.02$ps. The validation loss tracks the training loss meaning that the model is able to learn well. The testing loss also matches which means we are not over-fitting. The model is trained once on the synthetic data and used for inductive inference on a test set of open-source designs that it has not seen in the next section.

\subsection{Accuracy Evaluation}

In order to evaluate our model's ability to inductively infer results on unseen realistic designs, we evaluate on a subset of open-source designs created with the OpenROAD flow. In contrast to the MSE for training loss, the delay and slew Mean Absolute Error (MAE) from the GATMesh model predictions and first-order model~\cite{desai:meshsizing} are shown in Table~\ref{table:delay-slew-results}. We use SPICE simulation results as the reference to measure the error.  Against SPICE, our model performs competitively with the worst delay MAE being only $16.16$ps and, on average, only $5.27$ps. It also does quite well in predicting slews, with the worst slew MAE being $13.5$ps and, on average, $5.98$ps. The worst case design for both delay and slew is swerv\_wrapper which has significantly larger area compared to our training statistics. We could likely improve these results with a more representative training set. On the other hand, our model significantly outperforms the first-order model which has a worst mean delay error of $278.45$ps and, on average, $200.19$ps. The first-order model has no slew to compare and using an Elmore model approximation such as $2.2\times$ delay would be quite unfair and a poor comparison given that the delays are already quite inaccurate.

\subsection{Run-Time Evaluation}

\begin{table}[hbt] 
\caption{Runtime Analysis of GATMesh, First-Order Model~\cite{desai:meshsizing}, and Ngspice}
\begin{center}
\begin{tabular}{|c|r|r|r|} 
\hline
\textbf{Open-Source} &\multicolumn{3}{|c|}{\textbf{Runtime (s)}}\\
\cline{2-4} 
\textbf{Designs} & \textbf{\textit{GATMesh}} & \textbf{\cite{desai:meshsizing}} & \textbf{\textit{Ngspice}} \\
\hline
gcd & 0.0038 &  0.0003 & 0.2094 \\
\cline{1-4} 
aes & 0.0039 &  0.0527 & 2.5333 \\
\cline{1-4} 
ibex & 0.0041 &  0.0637 & 14.4586 \\
\cline{1-4} 
dynamic\_node & 0.0041 &  0.6546 & 16.8264 \\
\cline{1-4} 
tiny\_rocket  & 0.0044 & 2.4565 & 41.7799 \\
\cline{1-4} 
jpeg & 0.0046 &  3.5786 & 49.8608 \\
\hline
swerv & 0.0088 &  29.5014 & 851.3769 \\
\hline
swerv\_wrapper & 0.0087 &  28.3169 & 795.6460 \\
\cline{1-4}
\textbf{Average Runtime} & \textbf{0.0047} & \textbf{8.0780} & \textbf{221.5864} \\
\hline
\textbf{Average Speedup} & \textbf{1$\times$} & \textbf{1718$\times$} & \textbf{47146$\times$} \\
\hline
\end{tabular} \label{table:runtime}
\end{center}
\end{table}

We also present a runtime analysis with our GATMesh model inference, the first-order model algorithm, and Ngspice simulation~\cite{ngspice} in Table~\ref{table:runtime}. Ngspice, version 42, is run with the default two threads, but using more threads did not see any significant performance gain on these benchmarks. This is because the model evaluation of the mesh buffers can be easily parallelized, but the solution of the RC mesh equations cannot be parallelized very well at each simulation time step. The first-order model is faster than Ngspice but as the design sizes get bigger the runtime starts to increase even beyond that of our model. The prior work~\cite{desai:meshsizing} showed more than 24 hour run-times for the first-order algorithm on the largest designs. Overall, our model has a nearly constant runtime even for larger designs due to inference being faster with parallelization on GPUs. It is $1718\times$ faster on average than first-order model and on average $47146\times$ faster than Ngspice.

\subsection{Ablation Study}

We investigated the effectiveness of the different modeling decisions and how they impact the capability of the model to learn mesh timing. Specifically, we examine the static auxiliary connections that improve message passing from drivers to sinks and between buffers. We compare whether this static architecture is preferable to JK connections which learns to group features in different neighborhoods and between different levels of a multi-layer GNN. 

To do this, we did an ablation study of four different models. The first base case removes both the JK connections and the auxiliary connections in the graph but relies on the RC mesh and associated features. The features include derived ones such as regional capacitance, minimum resistance from nearby buffer, and total resistance from nearby buffers which do give some regional perspective of the RC mesh. These features, however, do not enable improvements in message passing like the auxiliary connections or learning over multiple layers like JK connections. They do, however, utilize multi-head attention for computing features from layer to layer.

The other models still utilize the same RC mesh and features, except that the second model adds the auxiliary connections but no JK connections, and the third model adds only JK connections but no auxiliary connections. The last is our GATMesh model with both auxiliary connections and JK connections. All the other parameters are the same as described in Table~\ref{table:hyperparameters}. 

\begin{figure}[htb]
    \centering
    \includegraphics[width=0.8\linewidth]{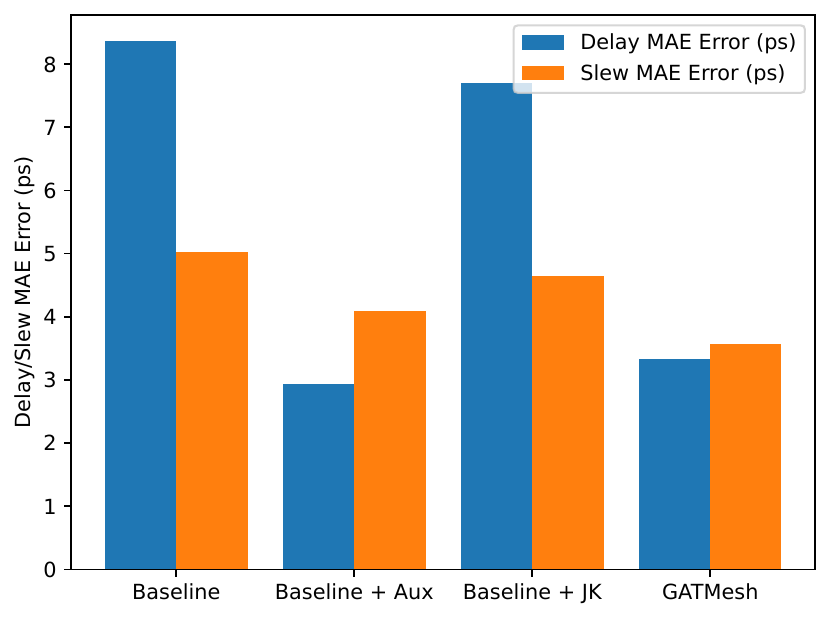}
    \caption{The addition of the auxiliary connections offers the most benefit for the delay MAE, however a combination of the auxiliary and JK connections proves better for slew.}
    \label{fig:ablation}
\end{figure}

We trained and tested each of these models on the identical synthetic dataset for comparison and present the results in Figure~\ref{fig:ablation}.  For both the delay and slew prediction, our GATMesh model produces substantial improvements over the baseline model without the static aux or learnable JK connections. The auxiliary connections alone produce significant gains on their own with the delay being slightly better than GATMesh but the slew being slightly worse. The JK connections alone achieve only moderate delay improvement, and almost no impact on slew error. 

\section{Conclusion}
We present GATMesh, a novel framework for accurate and efficient clock mesh timing analysis using Graph Neural Networks. By modeling the mesh as a graph of resistive-capacitive elements, including some derived features, and incorporating both learned (Jumping Knowledge connections) and domain-driven (auxiliary connections) enhancements, GATMesh effectively captures the complex interactions of multiple drivers, reconvergent paths, and spatial design variations. Our method models delay as well as slew of the signal and achieves predictions within a few picoseconds of SPICE accuracy while delivering massive speedups. Through extensive experiments and ablation studies, we show that both static and learnable architectural augmentations help prediction accuracy. GATMesh offers a practical and scalable path toward integrating clock meshes in real-world ASIC design flows by enabling faster and more accurate clock mesh analysis.

\clearpage
{\small
\bibliographystyle{ieeetr}
\bibliography{main}

\begin{thebibliography}{10}

\bibitem{desai:meshsizing}
M.~P. Desai, R.~Cvijetic, and J.~Jensen, ``Sizing of clock distribution networks for high performance {CPU} chips,'' in {\em Design Automation Conference (DAC)}, pp.~389--394, 1996.

\bibitem{power4_clock}
P.~Restle, C.~Carter, J.~Eckhardt, B.~Krauter, B.~McCredie, K.~Jenkins, A.~Weger, and A.~Mule, ``The clock distribution of the {POWER4} microprocessor,'' in {\em International Solid-State Circuits Conference (ISSCC)}, pp.~144--145, 2002.

\bibitem{power6_clock}
M.~G.-R. Thomson, P.~J. Restle, and N.~K. James, ``A {5GHz} duty-cycle correcting clock distribution network for the {POWER6} microprocessor,'' in {\em International Solid-State Circuits Conference (ISSCC)}, pp.~1522--1529, 2006.

\bibitem{amd:piledriver}
V.~Sathe, S.~Arekapudi, C.~Ouyang, M.~Papaefthymiou, A.~Ishii, and S.~Naffziger, ``Resonant clock design for a power-efficient high-volume x86–64 microprocessor,'' in {\em 2012 IEEE International Solid-State Circuits Conference}, pp.~68--70, 2012.

\bibitem{amd:bulldozer}
H.~McIntyre, S.~Arekapudi, E.~Busta, T.~Fischer, M.~Golden, A.~Horiuchi, T.~Meneghini, S.~Naffziger, and J.~Vinh, ``Design of the two-core x86-64 amd “bulldozer” module in 32 nm soi cmos,'' {\em IEEE Journal of Solid-State Circuits}, vol.~47, no.~1, pp.~164--176, 2012.

\bibitem{guthaus:highperf}
M.~Guthaus, X.~Hu, G.~Wilke, G.~Flache, and R.~Reis, ``High-performance clock mesh optimization,'' {\em ACM Transactions on Design Automation of Electronic Systems (TODAES)}, 2012.

\bibitem{meshworks}
A.~Rajaram and D.~Z. Pan, ``Meshworks: an efficient framework for planning, synthesis and optimization of clock mesh networks,'' in {\em Asia and South Pacific Design Automation Conference (ASP-DAC)}, pp.~250--257, 2008.

\bibitem{teng:mesh}
Y.~Teng and B.~Taskin, ``Clock mesh synthesis method using the earth mover's distance under transformations,'' in {\em 2012 IEEE 30th International Conference on Computer Design (ICCD)}, pp.~121--126, 2012.

\bibitem{wilke:buffer}
G.~Wilke, R.~Fonseca, C.~Mezzomo, and R.~Reis, ``A novel scheme to reduce short-circuit power in mesh-based clock architectures,'' in {\em Symposium on Integrated Circuits and System Design (SBCCI)}, pp.~117--122, 2008.

\bibitem{guthaus:mesh}
M.~R. Guthaus, G.~Wilke, and R.~Reis, ``Non-uniform clock mesh optimization with linear programming buffer insertion,'' in {\em Design Automation Conference (DAC)}, 2010.

\bibitem{friedman:mesh}
A.~Abdelhadi, R.~Ginosar, A.~Kolodny, and E.~G. Friedman, ``Timing-driven variation-aware nonuniform clock mesh synthesis,'' in {\em Great lakes Symposium on VLSI (GLSVLSI)}, pp.~15--20, 2010.

\bibitem{oversmoothingsurvey}
T.~K. Rusch, M.~M. Bronstein, and S.~Mishra, ``A survey on oversmoothing in graph neural networks,'' {\em arXiv:2303.10993}, 2023.

\bibitem{transformer}
A.~Vaswani, N.~Shazeer, N.~Parmar, J.~Uszkoreit, L.~Jones, A.~N. Gomez, L.~Kaiser, and I.~Polosukhin, ``Attention is all you need,'' in {\em Proceedings of the 31st International Conference on Neural Information Processing Systems}, NIPS'17, (Red Hook, NY, USA), p.~6000–6010, Curran Associates Inc., 2017.

\bibitem{jknet}
K.~Xu, C.~Li, Y.~Tian, T.~Sonobe, K.~ichi Kawarabayashi, and S.~Jegelka, ``Representation learning on graphs with jumping knowledge networks,'' {\em arXiv:1806.03536}, 2018.

\bibitem{bufformer}
R.~Liang, S.~Nath, A.~Rajaram, J.~Hu, and H.~Ren, ``Bufformer: A generative ml framework for scalable buffering,'' in {\em Proceedings of the 28th Asia and South Pacific Design Automation Conference}, ASPDAC '23, (New York, NY, USA), p.~264–270, Association for Computing Machinery, 2023.

\bibitem{Ye2023:WireTiming}
Y.~Ye, T.~Chen, Y.~Gao, H.~Yan, B.~Yu, and L.~Shi, ``Fast and accurate wire timing estimation based on graph learning,'' {\em Design, Automation \& Test in Europe Conference (DATE)}, 2023.

\bibitem{Cheng2020:WireTiming}
H.-H. Cheng, I.~H.-R. Jiang, and O.~Ou, ``Fast and accurate wire timing estimation on tree and non-tree net structures,'' {\em Design Automation Conference (DAC)}, 2020.

\bibitem{gat}
P.~Veličković, G.~Cucurull, A.~Casanova, A.~Romero, P.~Liò, and Y.~Bengio, ``{Graph Attention Networks},'' {\em arXiv:1710.10903}, 2018.

\bibitem{openroad}
A.~B. Kahng and T.~Spyrou, ``The {OpenROAD} project: Unleashing hardware innovation,'' in {\em Proc. GOMAC}, 2021.

\bibitem{boese:zeroskew}
K.~Boese and A.~Kahng, ``Zero-skew clock routing trees with minimum wirelength,'' in {\em ASIC Conf.}, pp.~1.1.1--1.1.5, 1992.

\bibitem{chao:zeroskew}
T.-H. Chao, Y.-C. Hsu, and J.~Ho, ``Zero skew clock net routing,'' in {\em Design Automation Conference (DAC)}, pp.~518--523, 1992.

\bibitem{tsai:clocksizing}
J.-L. Tsai, T.-H. Chen, and C.~C. Chen, ``Zero skew clock-tree optimization with buffer insertion/sizing and wire sizing,'' {\em {IEEE} Transactions on Computer-Aided Design of Integrated Circuits and Systems}, vol.~23, no.~4, pp.~565--573, 2004.

\bibitem{adam-opt}
D.~P. Kingma and J.~Ba, ``Adam: A method for stochastic optimization,'' 2017.

\bibitem{elu}
D.-A. Clevert, T.~Unterthiner, and S.~Hochreiter, ``Fast and accurate deep network learning by exponential linear units (elus),'' 2016.

\bibitem{pytorch}
J.~Ansel {\em et~al.}, ``{PyTorch 2: Faster Machine Learning Through Dynamic Python Bytecode Transformation and Graph Compilation},'' in {\em 29th ACM International Conference on Architectural Support for Programming Languages and Operating Systems, Volume 2 (ASPLOS '24)}, ACM, Apr. 2024.

\bibitem{pytorch_geometric}
M.~Fey and J.~E. Lenssen, ``Fast graph representation learning with {PyTorch Geometric},'' in {\em ICLR Workshop on Representation Learning on Graphs and Manifolds}, 2019.

\bibitem{ngspice}
{The NGSpice Project}, ``{ngspice -- open source circuit simulator, version 42}.'' \url{https://ngspice.sourceforge.io/}, 2024.
\newblock Accessed: 2025-04-12.

\end{thebibliography}
}
\end{document}